\documentclass[12pt]{article}
\usepackage{graphicx,subfigure}
\usepackage{latexsym,amsfonts,amssymb,epsfig,amsmath}

\newcommand{\be}{\begin{equation}}
\newcommand{\ee}{\end{equation}}
\newcommand{\beq}{\begin{eqnarray}}
\newcommand{\eeq}{\end{eqnarray}}
\newcommand{\beqn}{\begin{eqnarray*}}
\newcommand{\eeqn}{\end{eqnarray*}}
\newcommand{\ba}{\hspace*{-5pt}\begin{array}}
\newcommand{\ea}{\end{array}}

\newcommand{\bit}{\begin{itemize}}
\newcommand{\eit}{\end{itemize}}
\newcommand{\ben}{\begin{enumerate}}
\newcommand{\een}{\end{enumerate}}

\newtheorem{df}{Definition}

\newtheorem{prop}{Proposition}

\newtheorem{rem}{Remark}

\begin{document}

\begin{center}

{\bf\large Solitary waves in the model of active media, taking into account
 relaxing effects}


\vspace{8 mm}

{\it Wojciech Likus \footnote{e-mail: wojciech.likus@gmail.com} and Vsevolod A.~Vladimirov \footnote{e-mail: vsevolod.vladimirov@gmail.com}}

\vspace{8 mm}

{\it Faculty of Applied Mathematics,
AGH University of Science and Technology,\\
Mickiewicz Avenue 30, 30059 Krak\'{o}w, Poland}

\vspace{10 mm}

\end{center}

\noindent
{\bf Abstract.} {\small We study a system of differential equation simulating transport phenomena in active structured  media. The model is a generalization of  the McKean's modification of the celebrated FitzHugh-Nagumo system, describing the nerve impulse propagation in axon. It takes into account  the effects of memory, connected with the presence of internal structure. We construct explicitly the localized traveling wave solutions and analyze their stability.
}

\vspace{5mm}

\section{Introduction}
\label{introduction}
The second half of the XX-th century was marked by intense study of patterns' formation and evolution in open dissipative systems \cite{Prigogine}. A series of investigations was concerned with the problem of a nerve impulse formation and propagation along the neuron cells, the threshold effects and stability issues. One of the best known models describing the nerve impulse propagation in axon is that  put forward by Hodgkin and Huxley \cite{Hodgkin}. Being rather complicated for analytical treatment, the Hodgkin-Huxley model  was analyzed mainly by means of numerical methods. This circumstance inspired FitzHugh \cite{FitzHugh1,FitzHugh2} and independently Nagumo with co-workers \cite{Nagumo} for developing the following simplified model, maintaining the main features of the Hodgkin-Huxley equations:
\begin{eqnarray}
\frac{\partial\,u}{\partial\,t}=\frac{\partial^2\,u}{\partial\,x^2}-f(u)-w, \label{Fh_N1} \\
\frac{\partial\,w}{\partial\,t}=b\,u. \label{Fh_N2}
 \end{eqnarray}
 Here $f(u)=u\,(a-u)\,(1-u)$, and $a,\,\,b$ are positive constants. The system (\ref{Fh_N1})-(\ref{Fh_N2}) is used for the qualitative study of a nerve axon pulses, as well as for the description of general excitable media \cite{Winfree}. The nonlinearity of the function $f(u)$ in the FitzHugh-Nagumo model makes an obstacle for its analytical treatment and especially for obtaining exact solutions, so McKean \cite{McKean,Ping} introduces the following "caricature" of this system, still maintaining some qualitative features of the solutions describing the nerve impulses' propagation:
 \begin{eqnarray}
 \frac{\partial\,u}{\partial\,t}=\frac{\partial^2\,u}{\partial\,x^2}+H(u-a)-u-w, \label{McK1} \\
\frac{\partial\,w}{\partial\,t}=b\,u-d\,w, \label{McK2}
\end{eqnarray}
where $a,\,\,b$ are positive constants $d$ is nonnegative, and $H$ is the Heaviside function.
Rinzel and Keller \cite{RinzK} had constructed explicit solutions to these equations, describing the soliton-like traveling wave (TW) solutions and begun to study their stability. The stability investigations had been completed by J. Feroe, \cite{Feroe78}. Later on Klaasen and  Troy in \cite{Troy} and Evans, Fenich, Feroe in \cite{Feroe82} extended the study of solitary waves onto the more general systems of reaction-diffusion-kinetic equations, including terms, that enable to describe not only the nerve axon impulse creation and propagation, but many other patterns in dissipative systems, and in particular phenomena connected with the celebrated  Belousov-Zhabotinski reaction.

In this work we consider the following system:
\begin{eqnarray}
\tau\frac{\partial^2\,u}{\partial\,t^2}+\frac{\partial\,u}{\partial\,t}=\frac{\partial^2\,u}{\partial\,x^2}+H(u-a)-u-w, \label{McGen1} \\
\frac{\partial\,w}{\partial\,t}=b\,u-d\,w, \label{McGen2}
\end{eqnarray}
where $\tau\,\geq\,0$ is called the time of relaxation. A concept leading to the equation  with $\tau\,>\,0$ is presented in papers \cite{Joseph,Makar97,Kar03,DDMSV}. Eq. (\ref{McGen1}) can be formally introduced if one changes in the balance equation for the variable $u$ the conventional Fick's Law
\[
J(t,x)=-K \nabla \, Q(t,\,x),
\]
stating the generalized thermodynamical flow-force relation, with the Cattaneo's equation
\[
\tau\,\frac{\partial}{\partial\,t}\,J(t,\,x)+J(t,\,x)=-K \nabla \, Q(t,\,x),
\]
which takes into account the effects of memory connected with media internal structure.
Our goal is to construct the solitary wave solutions supported by the system (\ref{McGen1})-(\ref{McGen2}), to study their stability and impact of memory on the traveling waves. The paper is organized as follows. In Section 2 we construct exponentially localized generalized TW solutions to the system in question. In Section 3 we present the results of stability analysis of the solitary wave solutions. In the following section we briefly discuss the results obtained.


\section{Construction of the localized TW solutions}

We are looking for the TW solutions
 \[
 u(t,x)=u_c(z), \qquad w(t,x)=w_c(z), \qquad z=x+c\,t,
 \]
tending to zero as $|z|\rightarrow\infty$ .
Because of the invariance of the source system with respect to the  reflection $x\rightarrow-x$, it is sufficient to  focus on the case  $c>0$, corresponding to the TW moving to the left.

Substituting $u_{c}(z),\,\,w_{c}(z)$ into  (\ref{McGen1})-(\ref{McGen2}), we obtain the following system of ODEs:
\begin{equation}\label{TW}
\left\{\begin{matrix}
(\tau c^{2}-1)u''_{c}=-c u'_{c}- u_{c}+H(u_{c}-a)-w_{c}\\
cw_{c}'=b u_{c}-dw_{c}.
\end{matrix}\right.
\end{equation}
Let us assume that $a<\sup_{z\,\in\, \mathbb{R}}\, u_{c}(z)$, and the graph of the function  $u_{c}(z)$ intersects the constant function $g(z)=a$ only twice. Since the system (\ref{TW})  does not depend explicitly on $z$, we can make the invariant transformation $z\,\to z+b$, choosing $b$ in such a way that  $u_{c}(0)=a$ is the leftmost point of the intersection, while the other one is $z_1>0$.

 \begin{figure}
\begin{center}
\includegraphics[totalheight=1.8 in]{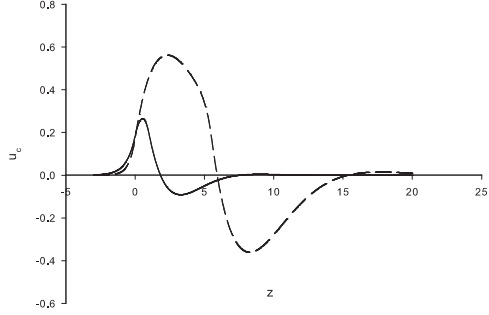}
\caption{The graphs of solutions $u_{c}$ obtained for $b=0.5,\ d=0.1,\ \tau=0.1$,  $c=c_{s}=0.65$ (solid line) and  $c=c_{f}=1.4$ (dotted line).}\end{center}
\end{figure}

The system (\ref{TW}) can be rewritten in the following expanded first-order form
\begin{equation}\label{OR}
\begin{bmatrix}
u_{c}'\\
u_{c}\\
w_{c}
\end{bmatrix}'=\begin{bmatrix}
\frac{c}{1-c^{2}\tau}& \frac{1}{1-c^{2}\tau} &\frac{1}{1-c^{2}\tau} \\
 1&0  &0 \\
 0&  \frac{b}{c}&\frac{-d}{c}
\end{bmatrix}
\begin{bmatrix}
u_{c}'\\
u_{c}\\
w_{c}
\end{bmatrix}-
\frac{1}{1-c^{2}\tau}H(u_{c}-a)\begin{bmatrix}
1\\
0\\
0
\end{bmatrix}.
\end{equation}
 The function $H[u_c(z)-a]$ is equal to unity on the interval $(0,z_{1})$ and nullifies elsewhere.  Therefore   on the set $\mathbb{R}\backslash (0,z_{1})$ the TW solutions satisfy the linear homogeneous system
\begin{equation}\label{lin}
V'=\hat A\,V,
\end{equation}
where $V=(u_{c}',u_{c},w_{c})^{T}$ and $\hat A$ is the matrix standing at the RHS of the system (\ref{OR}).
Let us introduce the notation
$\gamma={\left(1-c^{2}\tau\right)^{-1}}.
$  In this place we pose an additional condition $\tau\,c^2<1$, or $\gamma>0$, restricting from above possible values of the TW velocity.

It is easy to see that the matrix $\hat A$ has eigenvalues $\alpha_{1},\alpha_{2},\alpha_{3}$, satisfying the characteristic equation
\begin{equation}\label{charpol}
W(\alpha)=c \alpha^{3}+\alpha^{2}(d-c^{2}\gamma)-\alpha c\gamma(1+d)-\gamma(d+b)=0,
\end{equation}
while the corresponding eigenvectors are as follows:
\begin{equation*}
X_{j}=\begin{bmatrix}
\alpha_{j}\\
1\\
\frac{b}{c\alpha_{j}+d}
\end{bmatrix}.
\end{equation*}
Thus, any solution of the system (\ref{lin})  can be presented in the form
\begin{equation*}
V=\overset{3}{\underset{j=1}{\sum}} a_{j}e^{\alpha_{j} z}X_{i}.
\end{equation*}

 The characteristic polynomial $W(\alpha)$ has always one positive real root,  as it follows from the inequality $W(0)=-\gamma(d+b)<0$ and the asymptotic condition $\underset{\alpha\rightarrow +\infty}{lim}W(\alpha)=+\infty$.  Without loss of generality we can denote it by $\alpha_{3}$. Using the Viete formulae
\begin{eqnarray}
\alpha_{1}\alpha_{2}\alpha_{3}=\frac{\gamma(d+b)}{c}, \label{viete1} \\
\alpha_{1}\alpha_{2}+\alpha_{1}\alpha_{3}+\alpha_{2}\alpha_{3}=-\gamma(1+d),  \label{viete2} \\
\alpha_{1}+\alpha_{2}+\alpha_{3}=\frac{c^{2}\gamma-d}{c}, \label{viete3}
\end{eqnarray}
one can easily check that $Re(\alpha_{1}),\,\,Re(\alpha_{2})$ are negative.

On the interval $(0,z_{1})$, where the inhomogeneous term is nonzero, the general solution to (\ref{OR}) can be presented in the form
\begin{equation*}
V=\overset{3}{\underset{j=1}{\sum}} b_{j}e^{\alpha_{j} z}X_{j}+\hat A^{-1}
    \begin{bmatrix}
    \gamma \\
    0\\
    0
    \end{bmatrix}.
\end{equation*}
A solitary wave solution we are looking for vanishes as $z\to\,-\infty.$ Therefore  on the interval $\left(-\infty,\,\,0\right)$ the solution should be proportional to $\exp{(\alpha_3\,z)}\,X_3$. Adding the condition $u_{c}(0)=a$, we get
\begin{equation}\label{solit_z_minus}
    \begin{matrix}
    V(z)=e^{\alpha_{3}z}\begin{bmatrix}
    a\alpha_{3}\\
    a\\
    \frac{ab}{c\alpha_{3}+d}
    \end{bmatrix},&\ z<0.
\end{matrix}
\end{equation}
Taking advantage of the boundary conditions at the point $z=0,$  we obtain the extension of the solution (\ref{solit_z_minus}) onto the interval $(0,z_{1})$:
\begin{equation*}
\begin{matrix}
V(z)=M(z)M^{-1}(0)\left(\begin{bmatrix}
a\alpha_{3}\\
a\\
\frac{ab}{c\alpha_{3}+d}
\end{bmatrix}
-
\hat A^{-1}\begin{bmatrix}
 \gamma\\
0\\
0
\end{bmatrix}
\right)+
\hat A^{-1}
\begin{bmatrix}
 \gamma\\
0\\
0
\end{bmatrix}&\ \ \ 0< z<z_{1}
\end{matrix},
\end{equation*}
where $M(z)$ is the fundamental matrix of the system (\ref{lin}).
Similarly, taking advantage of the boundary conditions at the point $z=z_1,$ we obtain the extension of the solution onto the interval $(z_{1},+\infty)$:
\begin{equation}\label{auxforF}
\begin{matrix}
V(z)=M(z)M^{-1}(z_{1})V(z_{1}^{-})&\ \ \ z>z_{1}
\end{matrix}.
\end{equation}

In order that the condition $\lim\limits_{ z\to +\infty}u_c(z)=0$ be satisfied, the third coordinate of the vector
$
M^{-1}(z_{1})V(z_{1}^{-})
$
should be equal to zero.  This gives the condition
\begin{equation*}
\frac{\alpha_{1} \alpha_{2} \left(e^{\alpha_{3} z_{1}}-1\right) (d+\alpha_{3} c)-a (\alpha_{3}-\alpha_{1} ) (\alpha_{3}-\alpha_{2} ) e^{\alpha_{3}z_{1}} (b+d )}{(\alpha_{1}-\alpha_{3} ) (\alpha_{3}-\alpha_{2}) (b+d  )}=0,
\end{equation*}
or, which is the same,
\begin{equation}\label{warz1}
s:=e^{-\alpha_{3}z_{1}}=1-a \frac{(\alpha_{3}-\alpha_{1} ) (\alpha_{3}-\alpha_{2})\, (b+d )}{\alpha_{1} \alpha_{2} (d+\alpha_{3} c)}.
\end{equation}
The conditions $\alpha_{3}>0,\ z1>0$ immediately imply the inclusion $s \in\,(0,1)$.
Finally, taking into account the condition $u_{c}(z_{1})=a$, and putting accordingly the  second coordinate of $V(z_{1}^{+})$ in  (\ref{auxforF}) to be equal to $a$, we obtain:
\begin{eqnarray*}
F(a,b,c,d,\tau):=\frac{\alpha_{2}\alpha_{3} (d+\alpha_{1} c)\left(1-e^{\alpha_{1} z_{1}}\right)}{(\alpha_{1}-\alpha_{2}) (\alpha_{1}-\alpha_{3} ) (b+d)}+\\
\frac{\alpha_{1}\alpha_{3}  (\alpha_{2} c+d)\left(1-e^{\alpha_{2} z_{1}}\right)}{(\alpha_{2}-\alpha_{1}) (\alpha_{2}-\alpha_{3}) (b+d)}-a=0.
\end{eqnarray*}
Multiplication of $F(a,b,c,d,\tau)$ by
\begin{equation*}
\frac{(\alpha_{3}-\alpha_{1}) (\alpha_{3}-\alpha_{2}) (b+d)}{\alpha_{1} \alpha_{2} (d+c\alpha_{3})},
\end{equation*}
and employment of the equality  (\ref{warz1}) leads to the equation
\begin{equation*}
\begin{split}
h(s):=\left(\frac{\alpha_{2}\alpha_{3} (d+\alpha_{1} c)\left(1-s^{-\frac{\alpha_{1}}{\alpha_{3}} }\right)}{(\alpha_{1}-\alpha_{2}) (\alpha_{1}-\alpha_{3} ) (b+d)}+\frac{\alpha_{1}\alpha_{3}  (\alpha_{2} c+d)\left(1-s^{-\frac{\alpha_{2}}{\alpha_{3}}}\right)}{(\alpha_{2}-\alpha_{1}) (\alpha_{2}-\alpha_{3}) (b+d)}\right)\cdot \\ \cdot \frac{(\alpha_{3}-\alpha_{1}) (\alpha_{3}-\alpha_{2}) (b+d)}{\alpha_{1} \alpha_{2} (d+c\alpha_{3})}+s-1=0,
\end{split}
\end{equation*}
which, after some algebraic manipulation, can be presented in the form
\begin{equation}\label{ha_s}
\begin{split}
h(s)=\frac{-s^{-\frac{\alpha_{1}}{\alpha_{3}}}\alpha_{3}  (\alpha_{2}-\alpha_{3})(d+c\alpha_{1} )}
{\alpha_{1}(\alpha_{1}-\alpha_{2}) (d+c\alpha_{3}) }
+
\frac{-s^{-\frac{\alpha_{2}}{\alpha_{3}}}\alpha_{3} (\alpha_{3}-\alpha_{1})(d+c\alpha_{2})}
{\alpha_{2}(\alpha_{1}-\alpha_{2}) (d+c\alpha_{3})} +\\
+
\frac{\alpha_{3}  (\alpha_{2}-\alpha_{3})(d+c\alpha_{1})}
{\alpha_{1}(\alpha_{1}-\alpha_{2}) (d+c\alpha_{3}) }
+
\frac{\alpha_{3} (\alpha_{3}-\alpha_{1})(d+c\alpha_{2})}
{\alpha_{2}(\alpha_{1}-\alpha_{2}) (d+c\alpha_{3})}
+s-1=0.
\end{split}
\end{equation}
It is easily seen, that $
h(1)=h'(1)=0$. Hence, the existence of the local minimum of h(s) at the point 1 when $\underset{s\rightarrow 0^{+}}{lim}h(s)<0$ or the local maximum when $\underset{s\rightarrow 0^{+}}{lim}h(s)>0$, deliver the sufficient condition for the solvability of the equation $ h(s)=0$ inside the interval (0,1). As it is shown in \cite{RinzK}, these conditions are also necessary, assuring, in addition, the uniqueness of the solution.

\begin{figure}
\begin{center}
\includegraphics[totalheight=2.1 in]{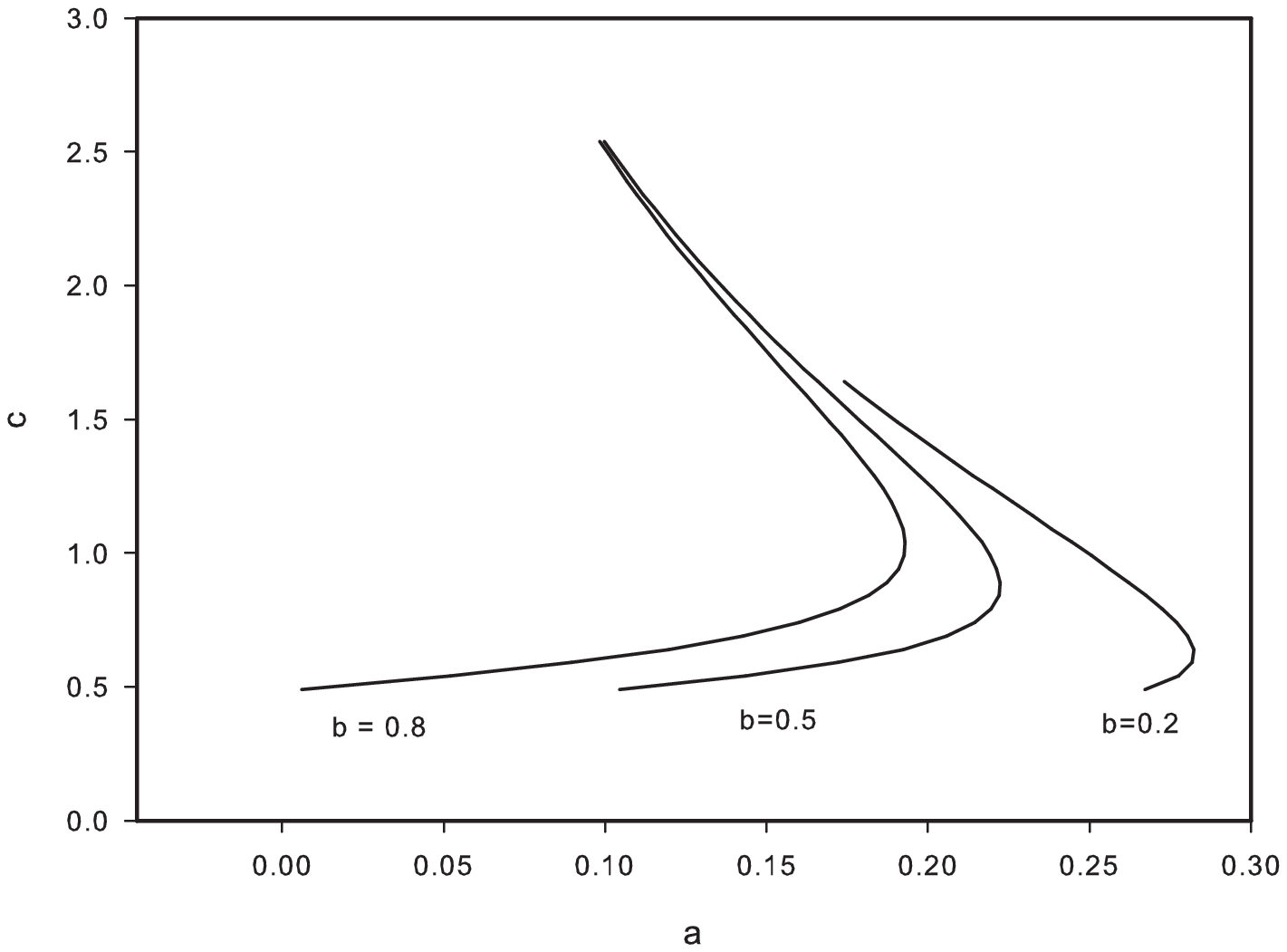}
\includegraphics[totalheight=2.1 in]{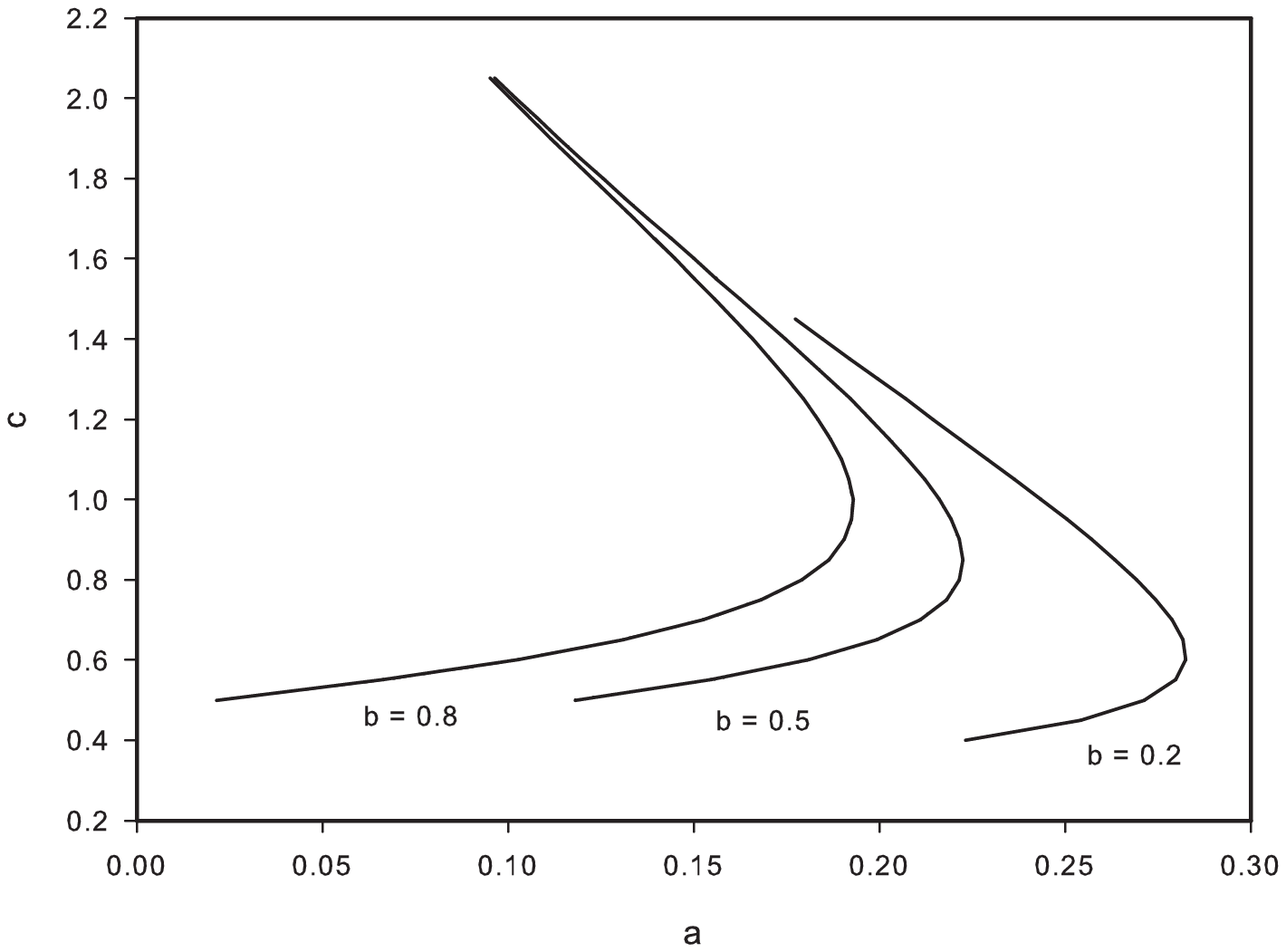}
\caption{Speed curves for different  values of $d,\ \tau$: left: $d=0.1,\ \tau=0.01$; right: $d=0.1,\ \tau=0.1$}\label{Fig:collisions}\label{KP1} \end{center}
\end{figure}

\begin{figure}
\begin{center}
\includegraphics[totalheight=2.1 in]{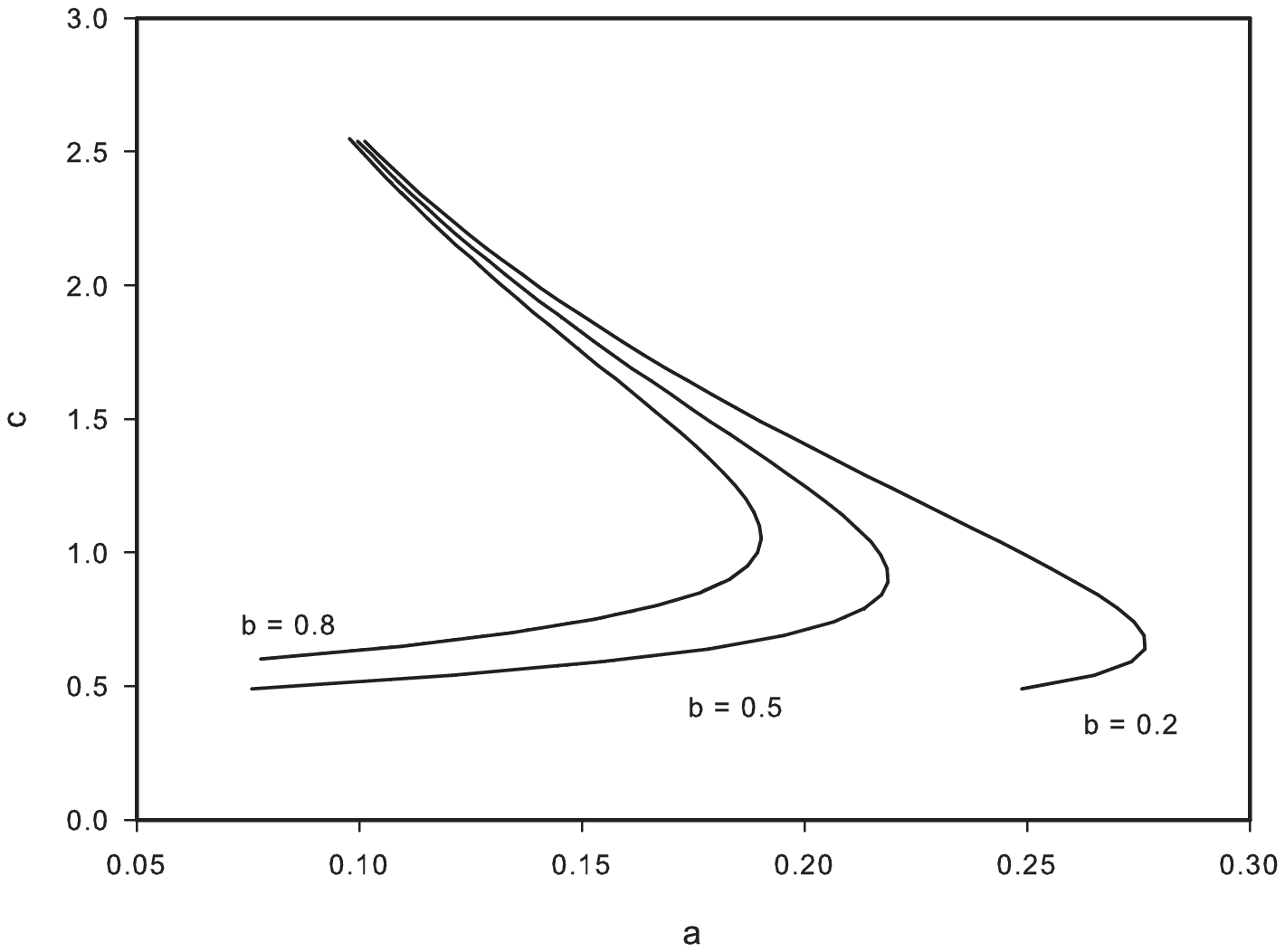}
\includegraphics[totalheight=2.1 in]{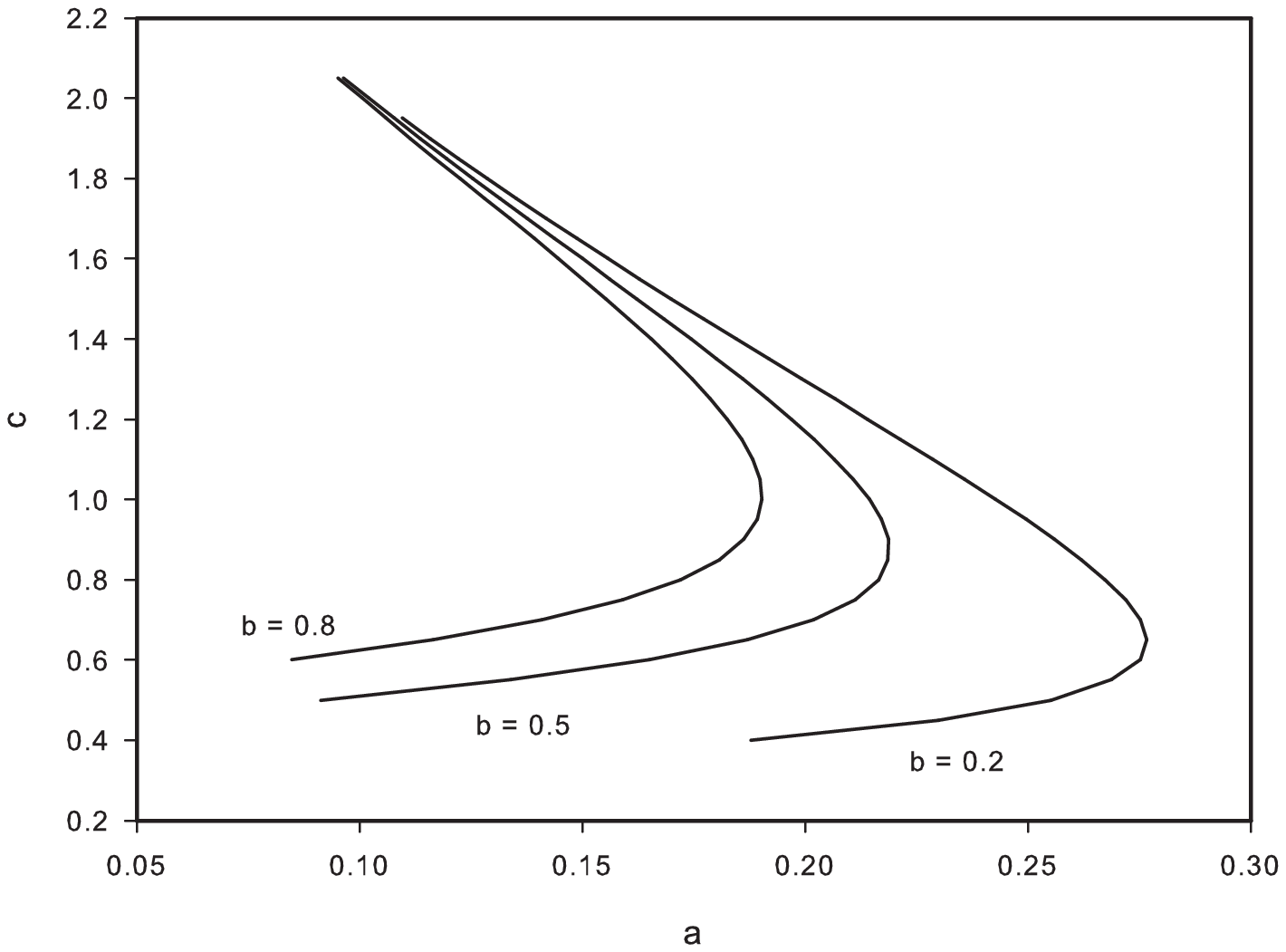}
\caption{Speed curves for different values of $d,\ \tau$: left: $d=0.05,\ \tau=0.01$; right: $d=0.05,\ \tau=0.1$}\label{Fig:collisions}\label{KP2} \end{center}
\end{figure}

For fixed $b,d$ and $\tau$, solution to the transcendental equation  $F(a,b,c,d,\tau)=0$ can be presented   as a graph of the function $c$ versus $a$, see Figs.~\ref{KP1}, \ref{KP2}. Note that on each graph presented there exists a point $a_{*}$, for which the  transcendental equation has a unique solution $c_{*}\left(a_{*};\,b,\,d,\,\tau\right)$. For $a\neq a_{*}$ the transcendental equation has two distinct solutions
$c_{s}\left(a;\,b,\,d,\,\tau\right)$ and $c_{f}\left(a;\,b,\,d,\,\tau\right)$, satisfying the inequalities $c_s<c_{*}<c_f$.
We can summarize the above construction in the form of the following statement.

\begin{prop}\label{stwh}
 Suppose that all the constants in the system (\ref{TW}) are positive, and following inequalities hold:
 \[c^2\,<1/\tau, \qquad d-d^2+2\,b>0. \]
 The system has  a unique exponentially localized solution if either
\begin{equation}\label{hmin}
W\left(\sqrt{\frac{\gamma(d+b)}{c^{2}\gamma+d}}\right)<0
<
W\left(\sqrt{\frac{\gamma(d-d^{2}+2b)}{d}}\right),
\end{equation}
or
\begin{equation}\label{hmax}
W\left(\sqrt{\frac{\gamma(d-d^{2}+2b)}{d}}\right)<0<W\left(\sqrt{\frac{\gamma(d+b)}{c^{2}\gamma+d}}\right).
\end{equation}
\end{prop}
Lengthy  technical proof of this statement is adduced in the Appendix A.

\section{Stability of the localized TW solutions}
\subsection{Statement of the problem}

In studying the spectral stability, it is instructive to pass to new  independent variables \[\bar{z}=x+ct, \qquad \bar{t}=t, \] in which the   TW solution $\left( u_{c},\,\,w_c \right)$ becomes stationary. In the new variables the system   (\ref{McGen1})-(\ref{McGen2}) reads as follows:
\begin{equation}\label{lineq}
\left\{\begin{matrix}
 \tau(c^{2}u_{\bar{z}\bar{z}}+2cu_{\bar{z}\bar{t}}+u_{\bar{t}\bar{t}})=
 u_{\bar{z}\bar{z}}-cu_{\bar{z}}-u_{\bar{t}}-u+H(u-a)-w, \\\\
 cw_{\bar{z}}+w_{\bar{t}}=bu-dw
\end{matrix}\right.
\end{equation}
(for the sake of simplicity, we will skip the lines above the independent variables). \\

We consider the perturbations of the following form:
\begin{equation}\label{stabrozw}
\begin{matrix}
u(z,t)=u_{c}(z)+\epsilon e^{\lambda t}U(z),
\\
w(z,t)=w_{c}(z)+\epsilon e^{\lambda t}W(z).
\end{matrix}
\end{equation}
Substituting  (\ref{stabrozw}) into the system (\ref{lineq})
and neglecting the $O(\epsilon^{2})$ terms,  we get a so called linearized (or variational) problem
\begin{equation}\label{variational}
\left\{\begin{matrix}
 \tau(c^{2}U''+2\lambda cU'+\lambda^{2}U)=U''-cU'-(\lambda +1)U-\frac{\delta(z-z_{1})}{u_{c}'(z_{1})}U+\frac{\delta(z)}{u_{c}'(0)}U-W, \\ \\
cW'+\lambda W=bU-dU.
\end{matrix}\right.
\end{equation}
which can be written as the first-order system
\begin{equation}\label{stabmac}
\begin{bmatrix}
U'\\
U\\
W
\end{bmatrix}'=
\begin{bmatrix}
\frac{c+2c\tau \lambda}{1- c^{2}\tau}&\frac{\tau \lambda^{2}+\lambda+1}{1- c^{2}\tau}&\frac{1}{1-c^{2}\tau}\\
1&0&0\\
0&\frac{b}{c}&\frac{-(d+\lambda)}{c}
\end{bmatrix}\begin{bmatrix}
U'\\
U\\
W
\end{bmatrix},
\end{equation}
endowed with the boundary conditions
\begin{equation}\label{vecbound}
\begin{matrix}
U'|^{^{0^{+}}}_{_{0^{-}}}=-\frac{U(0)}{(1-c^{2}\tau)u_{c}'(0)},&&
U'|^{^{z_{1}^{+}}}_{_{z_{1}^{-}}}=\frac{U(z_{1})}{(1-c^{2}\tau)u_{c}'(z_{1})}.
\end{matrix}
\end{equation}
\begin{df}
The set of all possible values $\lambda \,\in\,\mathbb{C}$ for which  (\ref{stabmac}), (\ref{vecbound}) has nontrivial bounded solutions is called the spectrum $\sigma$ of the linearized problem.
\end{df}

\noindent
We say that TW solution   $\left(u_{c}(z),\,w_{c}(z)\right)$  is spectrally stable if no
possible eigenvalue $\lambda$ for which the boundary value problem (\ref{stabmac}), (\ref{vecbound}) has a bounded nontrivial solution belongs to the right half-plane of the
complex plane.

\begin{rem}\label{lambzero} { It is not difficult to convince by the direct inspection that, differentiating the system (\ref{OR}) and making the substitution
$\left(u_c^{''},\,u_c^{'},\,w_c^{'}   \right)=\left(U^{'},\,U,\,W  \right),
$
we get a system of ODEs, coinciding with (\ref{stabmac}) for $\lambda=0$. In addition, the function $u_c^{''}(z)=U'(z)$ satisfies the jump conditions (\ref{vecbound}). From this we conclude that $0\,\in\,\sigma$.}
\end{rem}

Let us consider the characteristic polynomial of the matrix standing at the r.h.s. of (\ref{stabmac}):
\begin{eqnarray}
W_{\lambda}(\beta)=c\beta^{3}+(\lambda+d-c^{2}\gamma (1+2\lambda \tau))\beta^{2}-   \label{charlam}\\
-c \gamma(1+d+2d \tau \lambda  +3\lambda^{2}\tau+2\lambda)\beta- \gamma(\tau\lambda^{2}+\lambda+1)(\lambda+d)-b\gamma. \nonumber
\end{eqnarray}
We  denote the roots of the equation $W_{\lambda}(\beta)=0$  by $\left\{\beta_j  \right\}_{j=1}^3$.
 It is evident that an arbitrary solution to the  variational problem (\ref{stabmac}) takes the form of a linear combination of the vector-functions
\begin{equation*}
e^{\beta_{j}z}Y_{j}=e^{\beta_{j}z}
\begin{bmatrix}
\beta_{j}\\
1\\
\frac{b}{\beta_{j}c+d+\lambda}
\end{bmatrix},
\end{equation*}
where  $Y_{j}$ is the  eigenvector related to the eigenvalue $\beta_{j}$.  The real parts of the eigenvalues prove to satisfy some important relations described below.
\begin{prop}\label{wielchar}
For sufficiently small $|\lambda |$ the following relations hold true:
\begin{equation}\label{relpol}
Re(\beta_{3})>0,\qquad  Re(\beta_{1}) \,<0,\qquad Re(\beta_{2})<0.
\end{equation}
\end{prop}
{\bf Proof.} The characteristic polynomial is analytic  with respect to $\lambda$ and $\beta$ \cite{Bril}. That is why the change of signs of the real parts of  $\beta_{j}$  can occur only when one of the roots intersects the line
\[ \lambda(k)=\left\{ \lambda\,\in\,\mathbb{C}:\,\,W_{\lambda}(ik)=0\,\,k\,\in\,\mathbb{R}\right\}.\]
So the lines
$
\left\{W_{\lambda_{j}(k)}(ik)=0\right\}_{j=1}^3
$
divide the complex plane into the open sets in which the signs of  $\left\{Re(\beta_{j})\right\}_{j=1}^3$ remain constant.

Since the characteristic polynomial $W_{\lambda}(\beta)$, analytically depending on its coefficients,  coincides with (\ref{charpol}) for $\lambda=0$, then for sufficiently small $|\lambda|$, $\beta_{j}$ have the same character as $\alpha_{j}$ namely $Re(\beta_{3})>0$ and $Re(\beta_{1}),\,\,Re(\beta_{2})<0$.

\begin{rem}{ It is then evident that the signs of the real parts of these roots remain unchanged for any $\lambda\,\in\,\mathbb{C_{+}}$ if  $\left\{\lambda_{j}(k)\right\}_{j=1}^3$ lie in the left half-plane of the complex plane. Unfortunately, we do not have a suffice evidence on that $\lambda_j(k)\,\subset\,\mathbb{C_{-}}$, and we treat the inequalities (\ref{relpol})  as  conjectures that should be verified numerically (cf. with Figure~\ref{krzylam}).}
\end{rem}

\begin{figure}
\begin{center}
\includegraphics[totalheight=1.8 in]{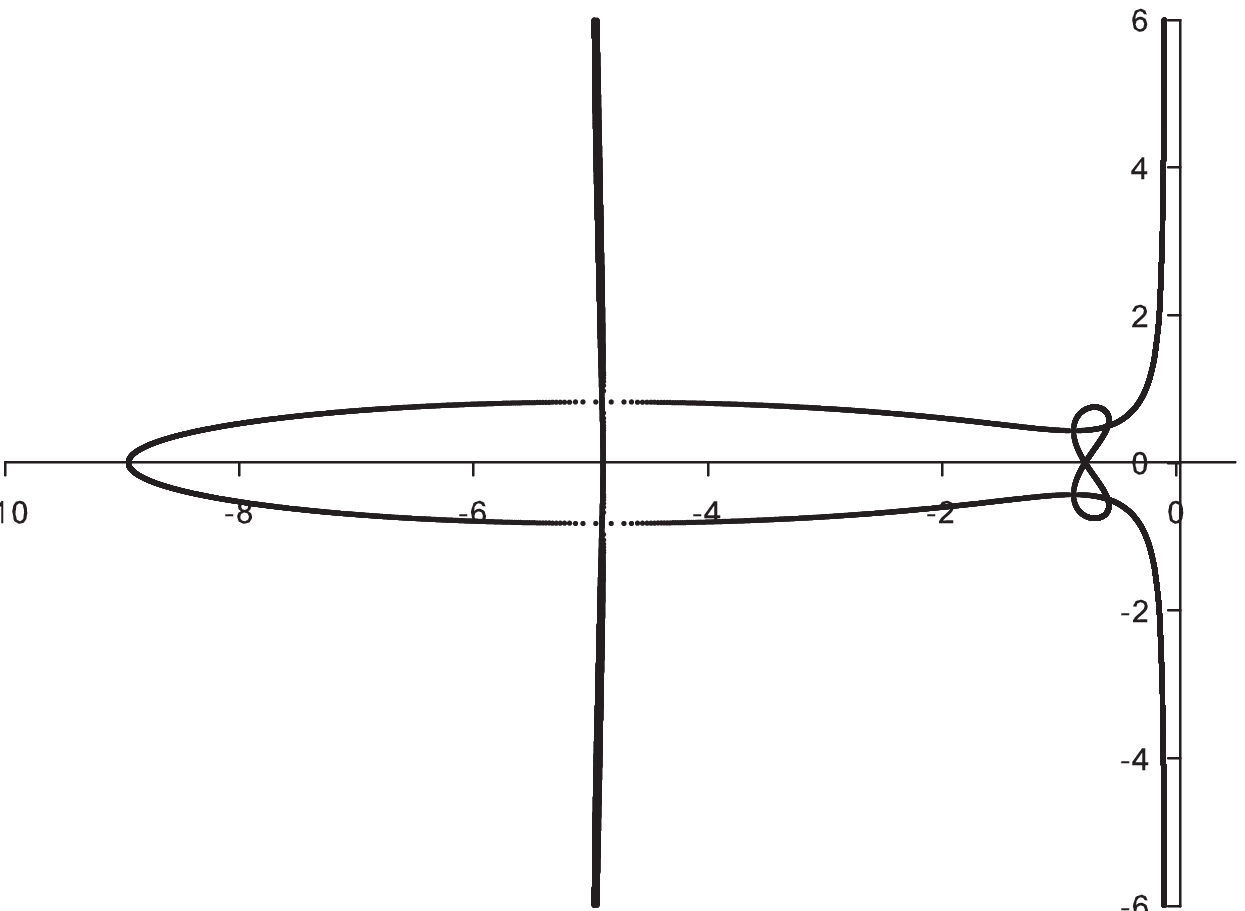}
\includegraphics[totalheight=1.8 in]{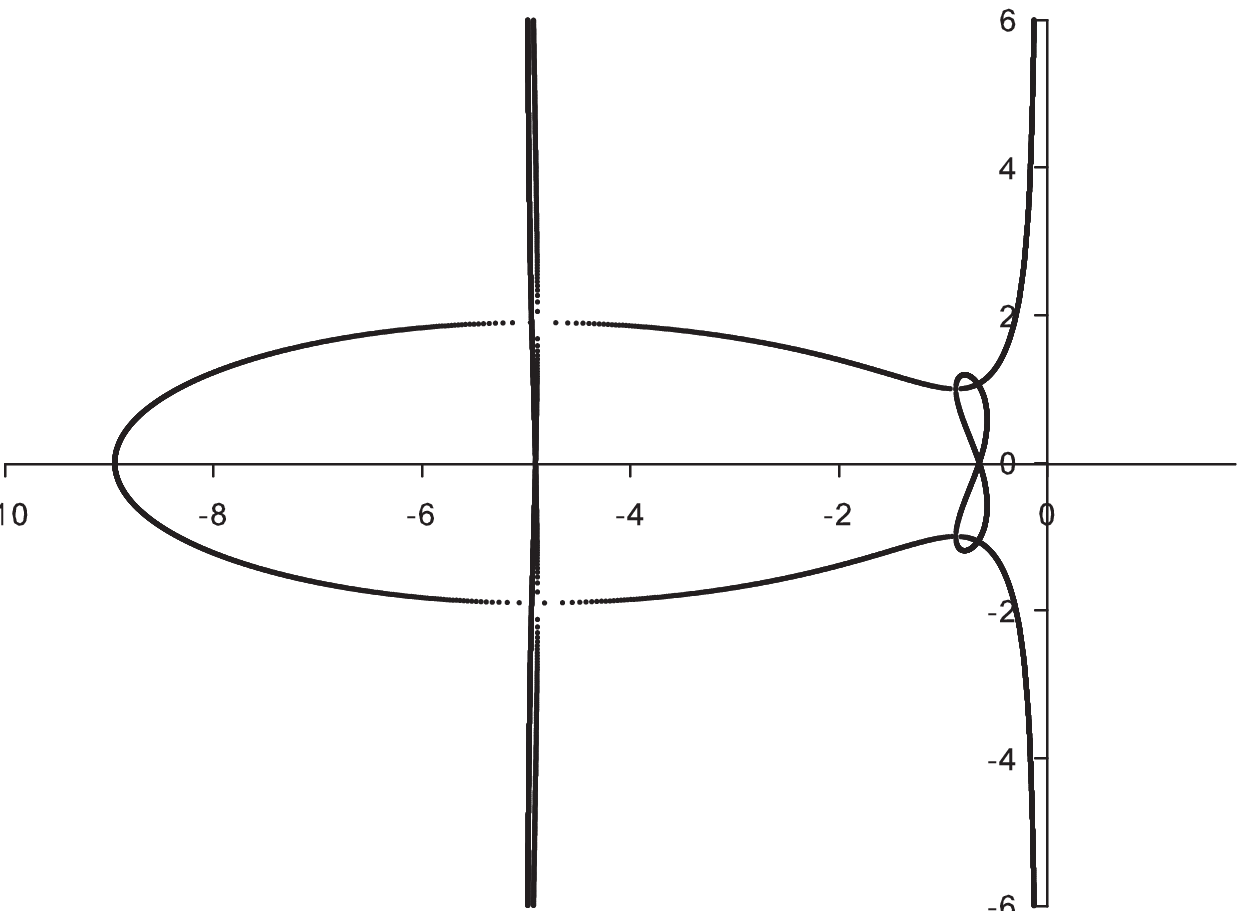}
\caption{Graphs of the parametrized curves $\lambda_j(k)$. Left:  $c_s = 0.6,\ d = 0.1,\ b = 0.3,\ \tau = 0.1$; right: $c_f = 1.5,\ d = 0.1,\ b = 0.5,\ \tau = 0.1.$ }\label{krzylam}\label{krzylam} \end{center}
\end{figure}

\subsection{Construction of the Evans function}

In studying the spectral stability of the TW solutions $\left(u_c(z),\,\,w_c(z)   \right)$, we  examine the existence of nontrivial  solutions to (\ref{stabmac}), (\ref{vecbound}) with $Re(\lambda)>0$,  treating  the inequalities (\ref{relpol}) as the only one possible arrangement of the roots of the characteristic polynomial $W_{\lambda}(\beta)$. Below we construct and analyze a so called Evans function, allowing for counting the number of eigenvalues with positive real parts, lying in a bounded region.

To begin with, let us note that under the above assumptions about $Re(\lambda_j)$,  any solution $Y^{-}(z,\lambda)$  vanishing at $-\infty$, is the element of subspace ${span}\{e^{\beta_{3}(\lambda)z}Y_{3}(\lambda)\}$. Solution to our problem  on the interval $(0,z_{1})$, satisfying the boundary condition at zero, takes the form
\begin{equation*}
J^{-}(z,\lambda)=M(z,\lambda)M(0,\lambda)^{-1}
\begin{bmatrix}
1&\frac{-1}{(1-c^{2}\tau)u'_{c}(0)}&0\\
0&1&0\\
0&0&1
\end{bmatrix}
Y^{-}(0^{-},\lambda),
\end{equation*}
where
\begin{equation*}
M(z,\lambda)=
\begin{bmatrix}
\beta_{1}(\lambda)e^{\beta_{1}(\lambda)z}&
\beta_{2}(\lambda)e^{\beta_{2}(\lambda)z}&
\beta_{3}(\lambda)e^{\beta_{3}(\lambda)z}\\\\
e^{\beta_{1}(\lambda)z}&
e^{\beta_{2}(\lambda)z}&
e^{\beta_{3}(\lambda)z}\\\\
\frac{b\ e^{\beta_{1}(\lambda)z}}{\beta_{1}(\lambda)c+d+\lambda}&
\frac{b\ e^{\beta_{2}(\lambda)z}}{\beta_{2}(\lambda)c+d+\lambda}&
\frac{b\ e^{\beta_{3}(\lambda)z}}{\beta_{3}(\lambda)c+d+\lambda}
\end{bmatrix}
\end{equation*}
is the fundamental matrix of Eq. (\ref{stabmac}).
Any solution $Y^{+}(z,\lambda)$ bounded on the interval $(z_1,\,+\infty)$, in turn, lies in the subspace  $span\{e^{\beta_{1}(\lambda)z}Y_{1}(\lambda),e^{\beta_{2}(\lambda)z}Y_{2}(\lambda)\}$.  Solutions lying in the interval $(0,z1)$,  and related with the solutions $\{Y_i^{+}\}_{i=1}^2$ via the boundary condition at $z=z_1$, can be presented in the form
\begin{equation*}
J^{+}_{1,2}(z,\lambda)=M(z,\lambda)M(z_{1},\lambda)^{-1}
\begin{bmatrix}
1&\frac{-1}{(1-c^{2}\tau)u'_{c}(z_{1})}&0\\
0&1&0\\
0&0&1
\end{bmatrix}
Y^{+}_{1,2}(z_{1}^{+},\lambda).
\end{equation*}
The function
\begin{equation}\label{Evans1}
\tilde E(\lambda):=Det(J^{-},J^{+}_{1},J^{+}_{2})(z_{1},\lambda)
\end{equation}
can be identified with the Evans function, widely used in literature \cite{Evans_3,Evans_4,Kapitula_Prom}. However, the function we are going to employ in order to reveal the spectrum location is somewhat different.
Note that the condition $\tilde E(\lambda)=0$ means that  $J^{-}(z_1,\lambda)$ is a linear combination of $J^{+}_{1}(z_1,\lambda), J^{+}_{2}(z_1,\lambda)$. This, in turn, gives the  possibility to extend the solution upon the interval $(z_1,\,+\infty)$, obtaining this way a solution defined on the real line and   vanishing at $\pm\,\infty$. Construction of such bounded solution will be possible then if after the extension of $J^{-}$ onto $(z_{1},+\infty)$ the third coefficient of decomposition on  the base  solutions $\left\{e^{\beta_{i}(\lambda)z}Y_{i}(\lambda)\right\}_{i=1}^3$, given by the formula
\begin{equation}\label{Evans2}
E(\lambda)=
\begin{bmatrix}
0&0&1
\end{bmatrix}
M(0,\lambda)^{-1}
\begin{bmatrix}
1&\frac{1}{(1-c^{2}\tau)u'_{c}(z_{1})}&0\\
0&1&0\\
0&0&1\\
\end{bmatrix}
J^{-}(z_{1},\lambda),
\end{equation}
will be equal to zero. The functions described by the formulae (\ref{Evans1}) and (\ref{Evans2}) occur to nullify simultaneously for $\lambda\,\in \,\mathbb{C}_{+}$.
Indeed, let us consider the function $\tilde E(\lambda)$.
 At  the point $z=z_{1}$
\begin{equation*}
J^{+}_{1,2}(z_{1},\lambda)=
\begin{bmatrix}
1&\frac{-1}{(1-c^{2}\tau)u'(z_{1})}&0\\
0&1&0\\
0&0&1
\end{bmatrix}Y_{1,2}(z^{+}_{1},\lambda)=BY_{1,2}(z^{+}_{1},\lambda).
\end{equation*}
Non-singularity of the matrix $B$ implies the equality
\begin{eqnarray*}
\tilde E(\lambda)=Det\left(J^{-}(z_{1},\lambda),BY_{1}(z_{1},\lambda),BY_{2}(z_{1},\lambda)\right)\\
=Det(B)Det\left(B^{-1}J^{-}(z_{1},\lambda),Y_{1}(z_{1},\lambda),Y_{2}(z_{1},\lambda)\right).
\end{eqnarray*}
The first column of the second matrix in the r.h.s. can be presented in the form of the following decomposition:
\begin{equation*}
B^{-1}J^{-}(z_{1},\lambda)=\underset{i=1}{\overset{3}{\sum}}b_{i}(z_{1},\lambda)Y_{i}(z_{1},\lambda).
\end{equation*}
Now, using the well-known properties of determinants,  we immediately get the formula:
\begin{eqnarray*}
\tilde E(\lambda)=Det(B)Det\left(b_{3}(z_{1},\lambda)Y_{3}(z_{1},\lambda),Y_{1}(z_{1},\lambda),Y_{2}(z_{1},\lambda)\right).
\end{eqnarray*}
The linear independence of $\left\{Y_j(z_1,\,\lambda)\right\}_{j=1}^3$, together with the non-singularity of the matrix $B$, implies that $\tilde E(\lambda)$ nullifies if and only if the
 $b_{3}(z_{1},\lambda)$ nullifies.
But
\begin{eqnarray*}
b_{3}(z_{1},\lambda)=[0,0,1]M^{-1}(z_{1},\lambda)B^{-1}J^{-}(z_{1},\lambda)=\\
=e^{-\beta_{3}z_{1}}[0,0,1]M^{-1}(0,\lambda)
\begin{bmatrix}
1&\frac{1}{(1-c^{2}\tau)u'(z_{1})}&0\\
0&1&0\\
0&0&1
\end{bmatrix}
J^{-}(z_{1},\lambda)=E(\lambda).
\end{eqnarray*}
The result obtained can be summarized as follows.

\begin{prop}
The function $\tilde E(\lambda)$ nullifies for $\lambda\,\in\,\mathbb{C}_{+}$ if and only if the function $E(\lambda)$ nullifies
\end{prop}

From now on we treat $E(\lambda)$ as the Evans function. Performing elementary algebraic manipulation, we can presented it in the following form:
\begin{equation}\label{Evans_expl}
E(\lambda)= U(0)+ \frac{\left( S_1+S_2 e^{-\beta_3 z_1}\right)\,(d+c\,\beta_3+\lambda)}{c\,(\beta_1-\beta_3)\,(\beta_2-\beta_3)},
\end{equation}
where
\begin{equation*}
S_1=\frac{-U(0)}{(1-c^{2}\tau)u'_{c}(0)},\qquad
S_2=\frac{U(z_{1})}{(1-c^{2}\tau)u'_{c}(z_{1})}.
\end{equation*}
It is worth noting that $\beta_{j}$ and $S_2$ depend on $\lambda$.
Now we formulate the crucial statement enabling to use the Evans function as a convenient
tools in studying the stability properties.

\begin{prop}\label{wlasevans} The following statements hold true:
\begin{enumerate}

\item $E(\lambda)$ is analytic function of $\lambda$,
\item $\overline{E(\lambda)}=E(\overline{\lambda})$,
\item $E(0)=0$,
\item $\underset{|\lambda|\rightarrow\infty}{lim}E(\lambda)=U(0)$.
\end{enumerate}
\end{prop}
The proof of this statement is supplemented in the Appendix B.

For the analytic function $E(\lambda)$  the following formula holds true \cite{Maurin}:
\begin{equation}\label{zeros_and_poles}
N=\frac{1}{2\pi i}\int_{\partial\,\Omega}\frac{E'(\lambda)}{E(\lambda)}d\lambda,
\end{equation}
where $N$  is  the total number of zeros (including multiplicities) contained in the bounded domain $\Omega$.
 Note, that $N$ is often referred to as {\it the winding number}, since it accounts the number of turns made by the parametrized curve $\left\{Re \left[E(\lambda)  \right],\,\,Im \left[E(\lambda)  \right]   \right\}$ around the origin as the argument runs along the closed curve $\partial\,\Omega$.

 Let us note \cite{Maurin}, that  $\partial\,\Omega$ should not contain zeros of the Evans function, therefore it  should not pass through the origin. The size of  the set $\Omega$ lying in the positive half-plane $\mathbb{C}_{+}$, should be sufficiently large in order to include all possible zeros of the Evans function. The smoothness of $E(\lambda)$ together with the asymptotics $\lim\limits_{|\lambda|\,\to\infty}E(\lambda)=U(0)$ guarantee  that the Evans function differs from zero when $|\lambda|$ is sufficiently large.

\begin{figure}
        \centering
                \includegraphics[totalheight=2.2 in]{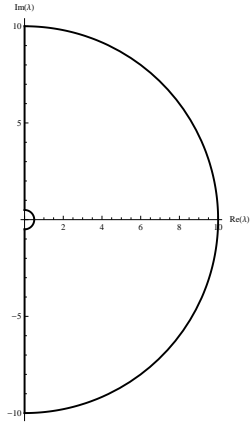}
                \label{fig1}
        \caption{A typical domain $\Omega$ is a  half-ring $P(r,\,R)$  of a radius $R>>1$   with clipped half-ring of a  radius  $r<<1$}\label{polpiers}
\end{figure}

We  begin analyzing the properties of the parametrized curves  $\left\{Re \left[E(\lambda)  \right],\,\,Im \left[E(\lambda)  \right]   \right\}$, called the Nyquist diagrams, for half-rings $\Omega=P(r,R)$ lying in $\mathbb{C}_{+}$ and separated from the origin, see Fig.~\ref{polpiers}.

\begin{figure}
\begin{center}
\includegraphics[totalheight=1.8 in]{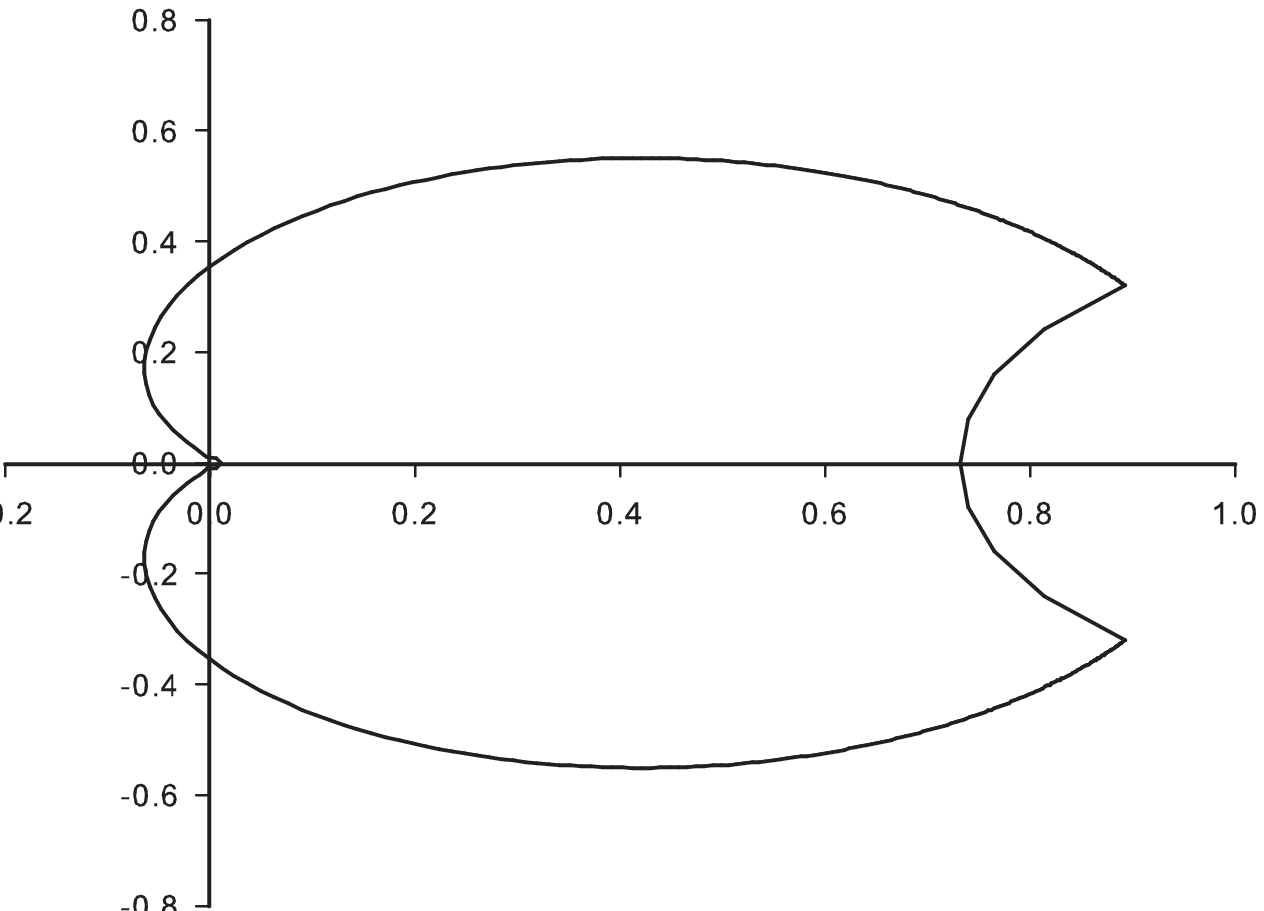}
\includegraphics[totalheight=1.4 in]{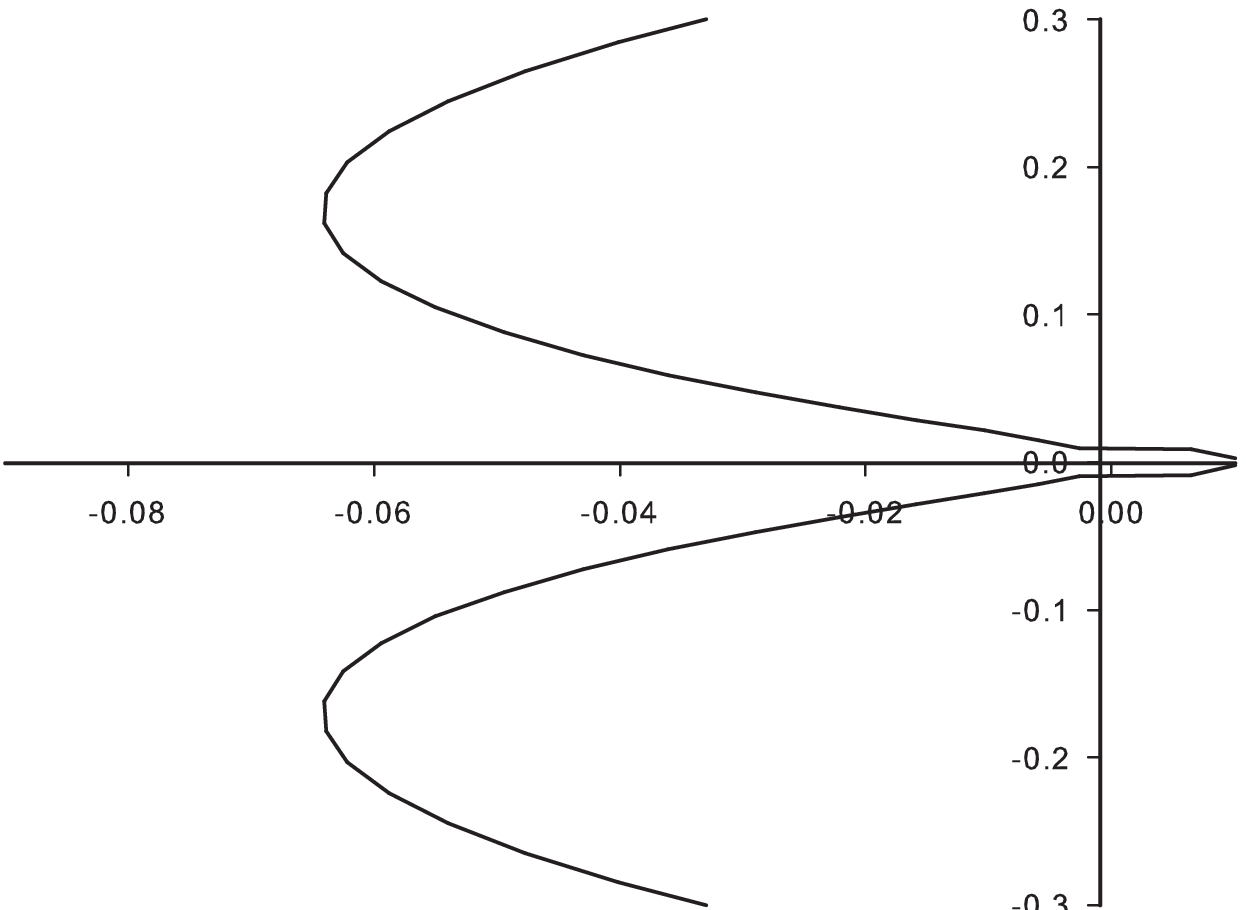}
\caption{The image of $ \partial P$ (left) and it's drawing near zero (right), obtained for the following values of the parameters: $b = 0.5,\, d = 0.1,\, \tau = 0.1$ and $c_{f} = 1.5, \, r=0.1,\,\,R=20$ }\label{PolEvans1} \end{center}
\end{figure}
\begin{figure}
\begin{center}
\includegraphics[totalheight=1.8 in]{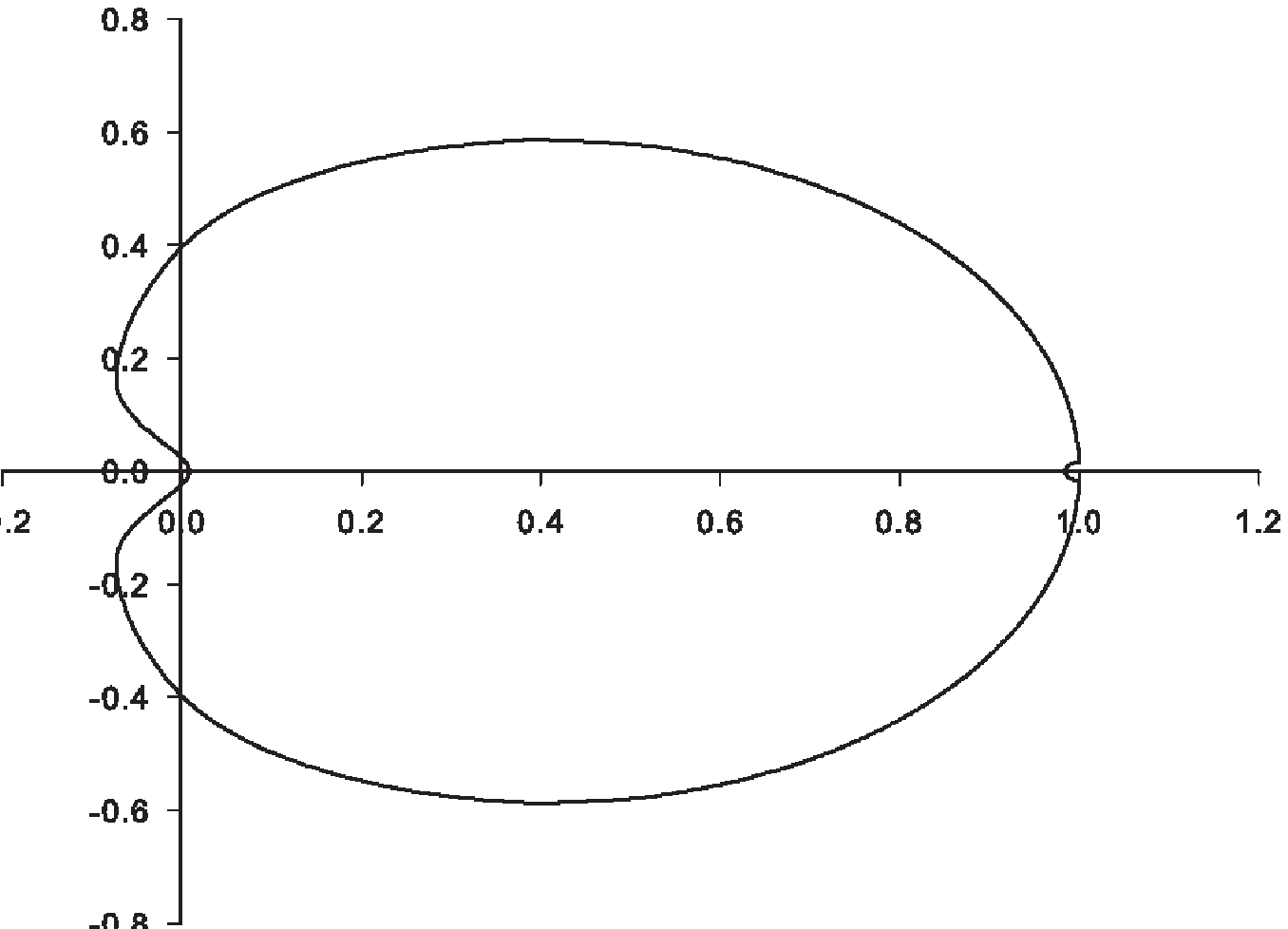}
\includegraphics[totalheight=1.4 in]{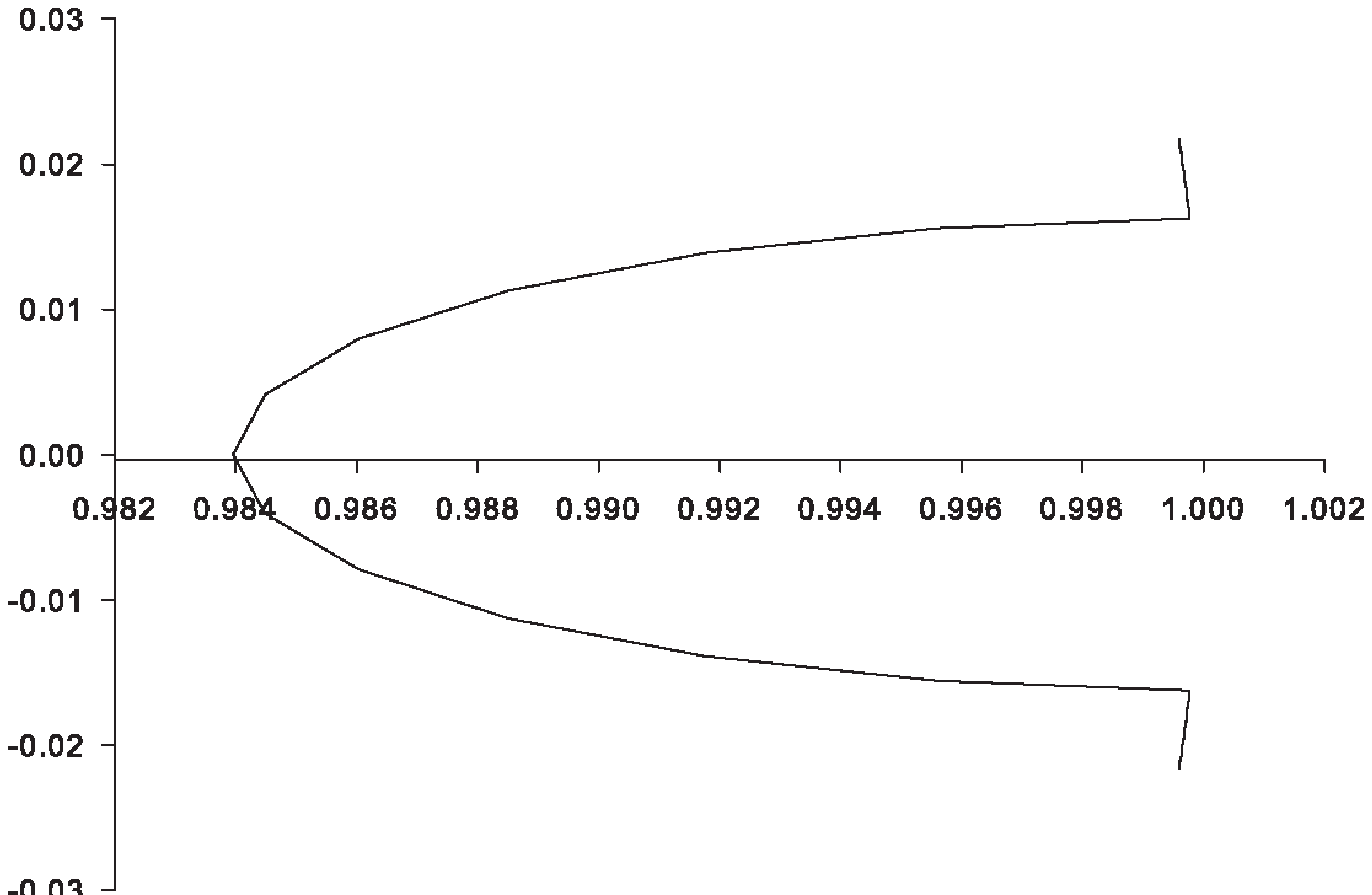}
\caption{The image of $ \partial P$ (left) and it's closeup near $U(0)=1$ (right) obtained for the following values of the parameters: $b = 0.5,\, d = 0.1,\, \tau = 0.1$ and $c_{f} = 1.5, \, r=0.2,\ R=1000$ }\label{PolEvans2} \end{center}
\end{figure}

The image of the Evans function calculated for $b = 0.5,\, d = 0.1,\, \tau = 0.1$, $r=0.1,\,\,R=20$ and $c_{f} = 1.5$ is shown on Fig.~\ref{PolEvans1}. We notice that the image does not wind around zero.
For sufficiently larger  radius $R=1000$   the winding number still is equal to zero (see Fig.~\ref{PolEvans2}).

\begin{figure}
\begin{center}
\includegraphics[totalheight=1.8 in]{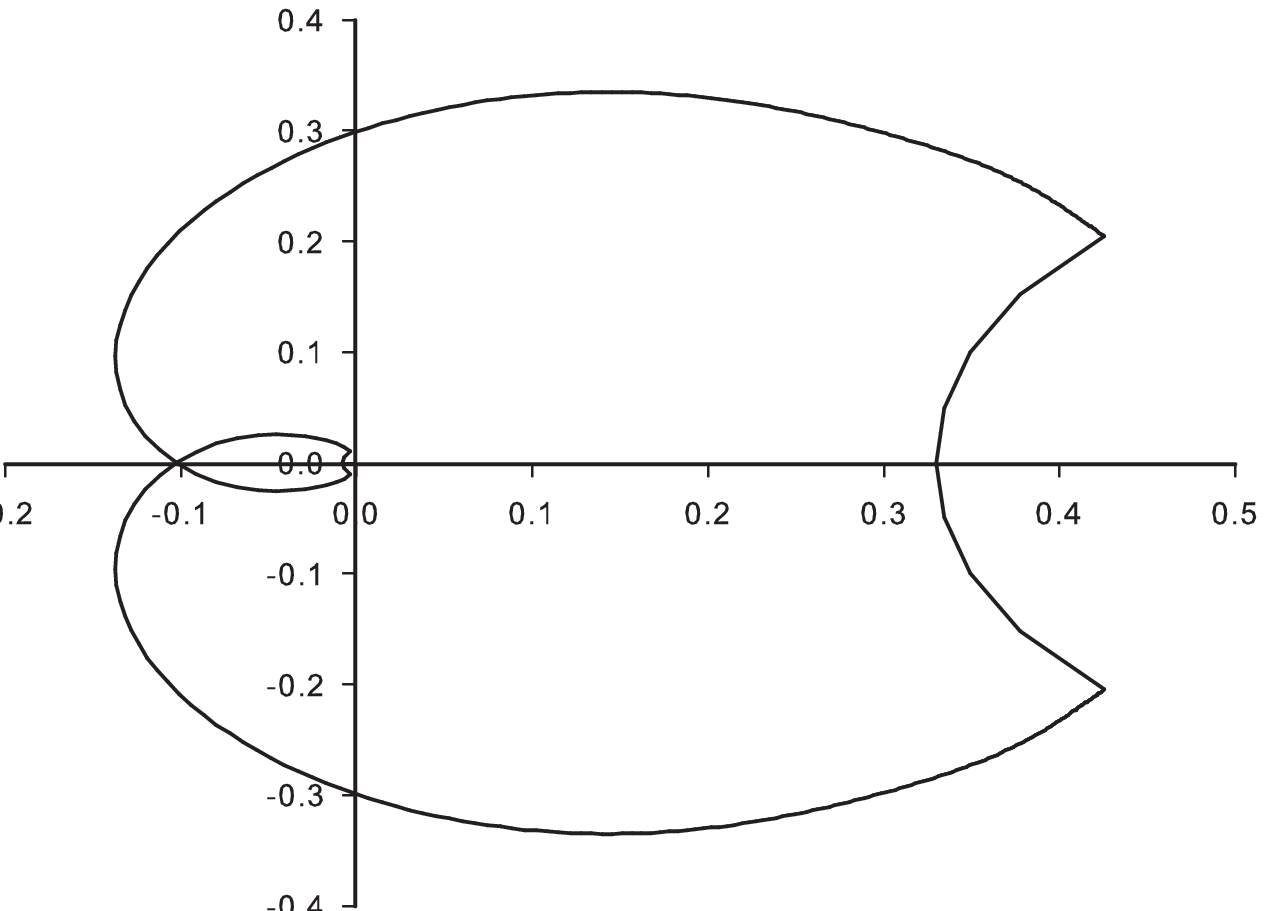}
\includegraphics[totalheight=1.4 in]{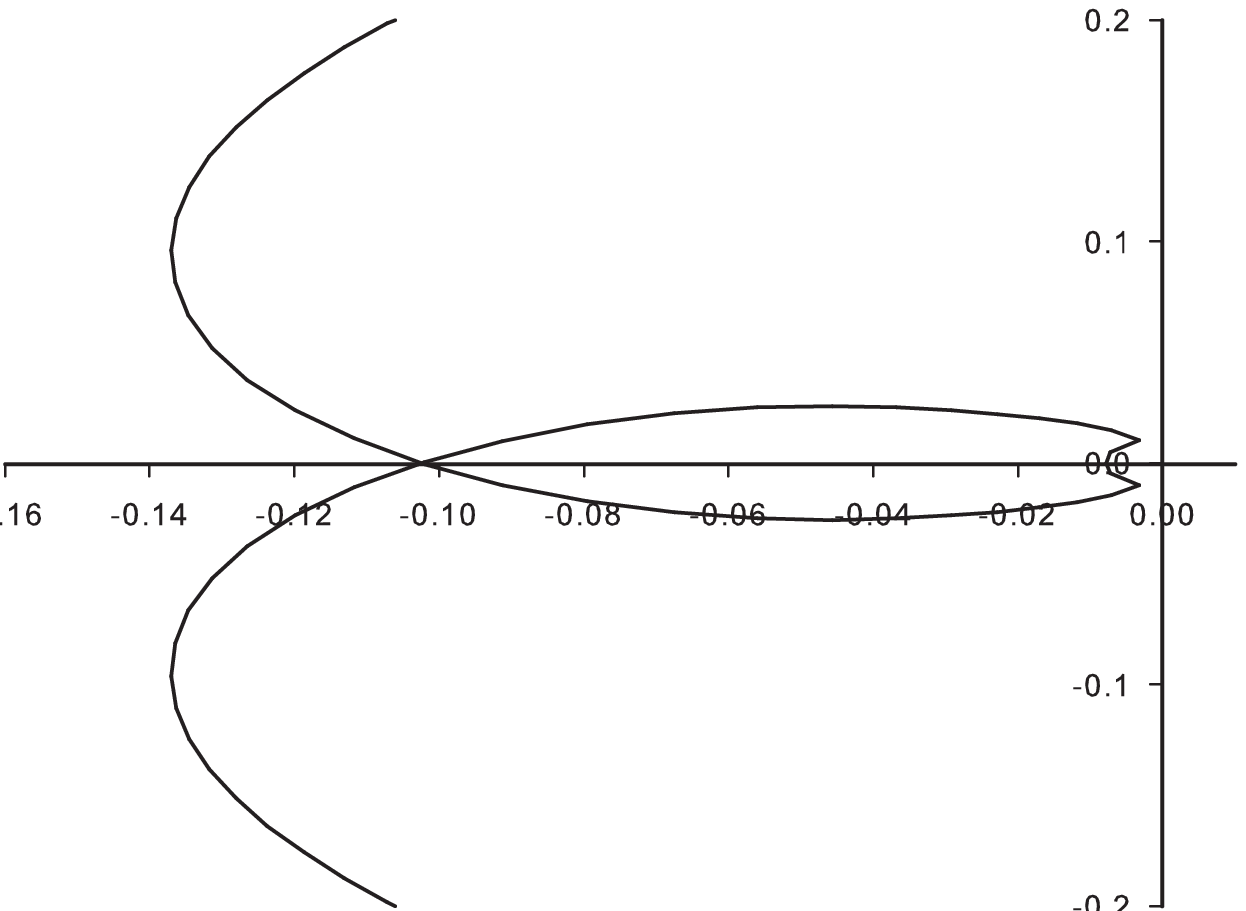}
\caption{The image of $ \partial P$ (left) and it's closeup near $U(0)=1$ (right) obtained for the following values of the parameters: $b = 0.5, \, d = 0.1,\ \tau = 0.1,\,r=0.1,\,R=20$ and the "slow" speed $c_s = 0.65$  }\label{PolEvans} \end{center}
\end{figure}

If we consider the image of the boundary of this half-rings obtained for $b = 0.5, \, d = 0.1,\ \tau = 0.1$ and the slower speed $c_s = 0.65,$ then we notice that the parametrized curve winds once around zero, as it is seen in Fig.~\ref{PolEvans}).

\begin{figure}
\begin{center}
\includegraphics[totalheight=1.6 in]{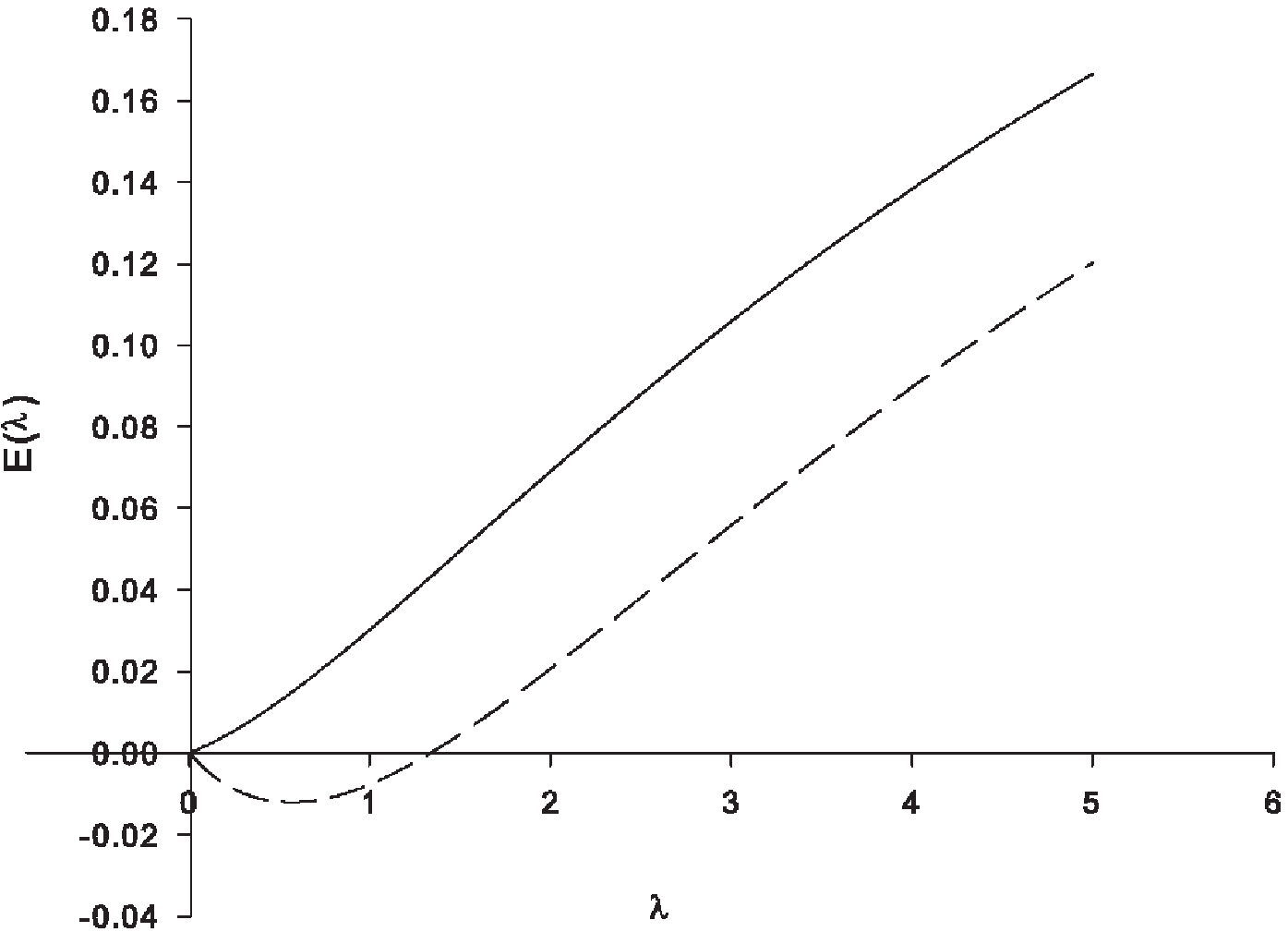}
\includegraphics[totalheight=1.6 in]{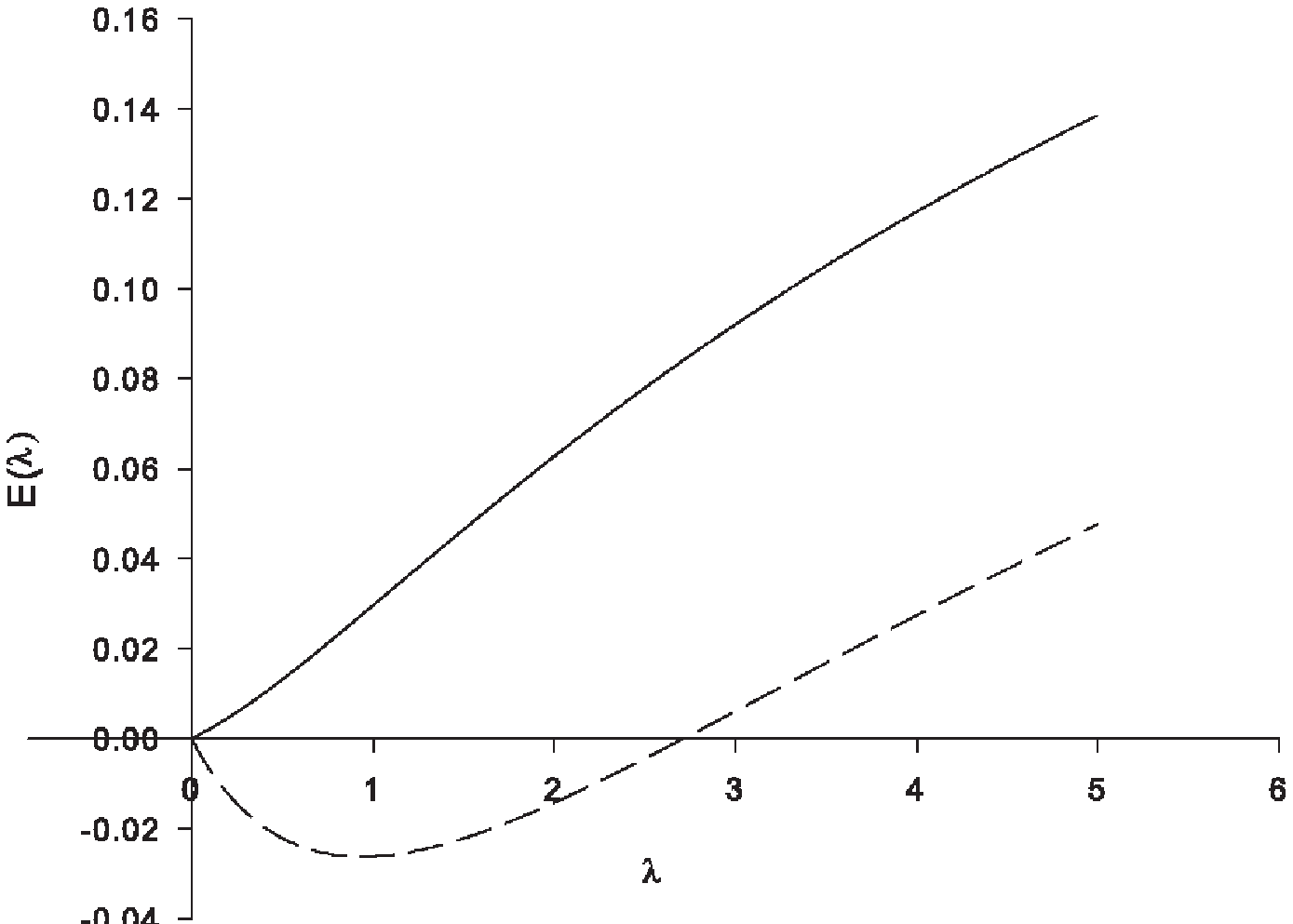}
\caption{The images of segments of $\mathbb{R}_{+}$ of $E(\lambda)$, for $d=0.1$, $b=0.5.$ Solid curves correspond to the "fast" speeds $c_{f}$, whereas dotted ones correspond to  the "lower" speeds $c_{s}$: left: $c_{f}=1.5,\,\,c_{s}=0.65,\ \tau=0.1$; right: $c_{f}=1.5,\,\,c_{s}=0.6,\ \tau=0.01$}\label{Evans} \end{center}
\end{figure}

To illustrate  the difference between the stable and unstable TW modes,  we plot the  images of a segment of the half-line $\mathbb{R}_{+}$ adjoining the origin for different values of the parameters, see Fig.~\ref{Evans}.
Presented plots confirm the conjecture about the instability of the "slow" modes, for which the Evans function has at least one zero in the right half-plane of the complex plane.

\section{Conclusions and discussion}

So in this work a model  the generalizing that of the McKean-FitsHugh-Nagumo and taking into account the memory effects has been considered. Let us present some conclusions appearing from the above investigations. First of all we notice that the properties of the exponentially localized TW obtained for $\tau>0$ are similar in many ways to the properties of the localized solutions supported by  the McKean model \cite{RinzK,Feroe78}, namely, at fixed values of the parameters $b,\,\tau$ and $d$ for any value of the parameter $a<a_{*}$ there is is a pair of values of the parameter, $c_s$ and $c_f$, obeying the inequality $c_s<c_f$. The "fast" solution, corresponding to $c_f$, is stable, while the "slow" one, corresponding to $c_s$, is unstable. Yet in contrast to the results concerning the McKean model, there is the  restriction $|c|<c_{cr}=1/\sqrt{\tau}$ for possible values of the TW velocity.  Our preliminary investigations show that there is no exponentially localized TW solutions for $|c|>c_{cr}$ in case we maintain  the RHS of the system (\ref{McGen1})-(\ref{McGen2}).

Let us notice in conclusion that numerous one-component models without kinetics, reminding Eq. (\ref{McGen1}), are investigated in \cite{VladKu_04,VladKu_05,VladKu_06,Vlad2008}, where a number of solitary wave solutions have been described. Yet the analytical study as well as the numerical investigations show that none of the solitary wave solution is stable \cite{VlaMacz_12} in absence of kinetic equations.


\section*{Appendix A.}
Proof of the proposition~\ref{stwh}.
We restrict ourselves to the case of the local minimum, for the proof for the local maximum is analogous. So we  wish to pose the conditions which guarantee that   $h(0^+)<0$ and $h''(1)>0$. To begin with, let us remark, that   for any $y \in \mathbb{C}$ such that $Re(y)>0$
$
\underset{s\rightarrow 0^{+}}{lim}s^{y}=0.
$
Returning to the formula (\ref{ha_s}) and taking into account that both $Re\left(-\frac{\alpha_{1}}{\alpha_{3}}\right)$ and $Re\left(-\frac{\alpha_{2}}{\alpha_{3}}\right)$ are positive we get:
\begin{equation*}
\underset{s\rightarrow 0^{+}}{lim}h(s)=
\frac{d(-\alpha_{3} (\alpha_{2}+\alpha_{1} )+\alpha_{3}^{2}+\alpha_{1}\alpha_{2})}
{\alpha_{1}\alpha_{2} (d+c\alpha_{3}) }
-2.
\end{equation*}
Multiplying the inequality
\[
\frac{d[-\alpha_{3} (\alpha_{2}+\alpha_{1} )+\alpha_{3}^{2}+\alpha_{1}\alpha_{2}]}
{\alpha_{1}\alpha_{2} (d+c\alpha_{3}) }
-2<0
\]
 by $\alpha_{1}\alpha_{2} (d+c\alpha_{3})>0$, we obtain the expression
\begin{equation*}
d\,[\alpha_{3}^{2}-(\alpha_{3} \alpha_{2}+\alpha_{3} \alpha_{1}+\alpha_{2} \alpha_{1})  ]
<2c\alpha_{1}\alpha_{2} \alpha_{3}.
\end{equation*}
Using the Viete formulae (\ref{viete1})-(\ref{viete3}), we can rewrite it as
\begin{equation*}
d(\gamma(1+d)  +\alpha_{3}^{2})
<2(d+b)\gamma,
\end{equation*}
and this finally gives us the inequality
\begin{equation}\label{nier1}
\alpha_{3}
<\sqrt{\frac{\gamma(d-d^{2}+2b)}{d}}.
\end{equation}

Now let us address the inequality  $h''(1)>0$. Calculating the second derivative of (\ref{ha_s}) and inserting $s=1$, we obtain:
\begin{equation*}
h''(1)=1+
\frac{d\alpha_{3}+c \alpha_{3}(\alpha_{1}+\alpha_{2}) -c\alpha_{1}\alpha_{2}}
{\alpha_{3} (d+c\alpha_{3})} >0.
\end{equation*}
Multiplying this inequality by $\alpha_{3} (d+c\alpha_{3})>0$ we get:
\begin{equation*}
2d\alpha_{3}+c \alpha_{3}(\alpha_{1}+\alpha_{2}) -c\alpha_{1}\alpha_{2}
+c\alpha_{3}^{2} >0.
\end{equation*}
Using the Viete formulae (\ref{viete2}) and (\ref{viete3})  we can rewrite it as
\begin{equation*}
2d\alpha_{3}+2c \alpha_{3}\left(\frac{c^{2}\gamma-d}{c}-\alpha_{3}\right) +c\gamma (1+d)
+c\alpha_{3}^{2} >0.
\end{equation*}
Multiplying the above inequality by $\alpha_3$, and next employing  the equation $W(\alpha_3)=0,$ we finally get, after some algebraic manipulation, the following inequality:
\begin{equation*}
\alpha_{3}^{2}>\frac{\gamma(b+d)}{d+c^{2}\gamma},
\end{equation*}
or, which is the same,
\begin{equation}\label{nier2}
\alpha_{3}>\sqrt{\frac{\gamma(b+d)}{d+c^{2}\gamma}}.
\end{equation}
The conditions (\ref{hmin}) immediately appear from the fact that the relations $0<z<\alpha_3<y$ imply the inequalities $W(z)<0<W(y)$.

\section*{Appendix B.}
Proof of the proposition~\ref{wlasevans}.

Ad 1. $E(\lambda)$ is a rational function of $\lambda$, moreover,  its denominator is strictly positive when $\lambda\,\in\,C_+\bigcup\{0\}$. And since both the numerator and denominator are superpositions of analytic function, then $E(\lambda)$ is also analytic.\\\\
Ad 2. The roots of $W_{\bar{\lambda}}(\beta)$ are complex conjugations of $\beta_{j}$. Using this fact, we conclude that  $E(\bar{\lambda})=\overline{E(\lambda)}$.\\\\
Ad 3. It is a simple consequence of the fact that $0\,\in\,\sigma$, see remark~\ref{lambzero}.\\\\
Ad 4. Dividing the characteristic polynomial $W_{\lambda}(\beta)$ by $\lambda^{3}$ we  get an expression of the form
\begin{equation*}
W_{\lambda}\left(\frac{\lambda}{\beta}\right)=O(1)\left(\frac{\beta}{\lambda}\right)^{3}+O(1)\left(\frac{\beta}{\lambda}\right)^{2}+
O(1)\left(\frac{\beta}{\lambda}\right)+O(1).
\end{equation*}
From this we conclude that
\begin{equation*}
\beta_{j}=\textsl{O}(|\lambda|), \qquad j=1,2,3.
\end{equation*}
Applying this, we obtain the following estimation for the derivative:
\begin{equation*}
\frac{dW_{\lambda}}{d \beta}(\beta_{j})=\textsl{O}(|\lambda|^{2}), \qquad j=1,2,3.
\end{equation*}
This gives us the following estimation for the Evans function's denominator:
\begin{equation*}
\left(\frac{1}{1-c^{2}\tau}\right)\frac{dW_{\lambda}}{d \beta}(\beta_{3})=c(\beta_{3}-\beta_{1})(\beta_{3}-\beta_{2})=\textsl{O}(|\lambda|^{2}).
\end{equation*}
We also use the estimations
\begin{equation*}
S_1=\textsl{O}(1),
\end{equation*}
and
\begin{align*}
S_2=\frac{e^{\beta_{3}z_{1}} (c (\beta_{3}S_1+U(0) (\beta_{1}-\beta_{3}) (\beta_{2}-\beta_{3}))-d S_1+\lambda  S_1)}{(\beta_{3}-\beta_{1}) (\beta_{3}-\beta_{2})}+\\
+\frac{S_1 e^{\beta_{1}z_{1}} (\beta_{1} c-d+\lambda )}{(\beta_{1}-\beta_{2}) (\beta_{1}-\beta_{3})}+\frac{S_1 e^{\beta_{2} z_{1}} (\beta_{2} c-d+\lambda )}{(\beta_{2}-\beta_{1}) (\beta_{2}-\beta_{3})}=\textsl{O}(U(0)e^{\beta_{3} z_{1}}).
\end{align*}
Now it is evident that
\begin{equation*}
\underset{|\lambda|\rightarrow+\infty}{lim}E(\lambda)=U(0).
\end{equation*}

\section*{ Acknowledgments.} The authors gratefully acknowledge support from the Polish Ministry
of Science and Higher Education.


\begin{thebibliography}{88}
\bibitem{Prigogine}
P. Glansdorff and Prigogine, I.,
{\em Thermodynamics of
Structure, Stability and Fluctuations},
Wiley-Interscience, New York, 1971.
\bibitem{Hodgkin}
A.L. Hodgkin and A.F. Huxley, {\em  J. Physics,} {\bf 177} 500 (1952).
\bibitem{FitzHugh1}
R. FitzHugh, {\em  Biophysical Journal,} {\bf 1} 445 (1969).
\bibitem{FitzHugh2}
R. FitzHugh, {\em Mathematical models of excitation and propagation in nerve,} in Biological Engineering. H.P. Schwan, ed., McGraw-Hill, Inc., NewYork, 1969.
\bibitem{Nagumo}
J. Nagumo, S. Arimoto and S. Yoshizawa, {\em  Proc. IRE,} {\bf 50} 2061 (1962).
\bibitem{Winfree}
A. Winfree, {\em Stately rotating patterns of reaction and diffusion,} in Progress in Theoretical Chemistry, Vol. 4, H. Eyring and D. Henderson eds., Academic Press, New York, 1978, pp. 1-51.
\bibitem{McKean} H.P. McKean, {\em Adv. Math.} {\bf 4} 209 (1971).
\bibitem{Ping} W.P. Wang {\em  Comm. on Pure and App. Math.,} {\bf XLI}  997 (1988).
\bibitem{RinzK} J.Rinzel , J.B.Keller, {\em  Biophysical Journal,} {\bf 13} 1313 (1973).
\bibitem{Feroe78}J. Feroe, {
Biophysical Journal} {\bf  21} 103 (1978).
\bibitem{Troy}
A. Klaasen and W. Troy, {\em   SIAM Journal of Appl. Math.,} {\bf 41} 145 (1981).
\bibitem{Feroe82}
J. Evans, N. Fenichel and J. Feroe, {\em  SIAM Journal of Appl. Math.,} {\bf  42} 219 (1982).
\bibitem{Joseph}
D.D. Joseph and L. Preziozi, {\em  Rev. Mod. Phys.,} {\bf  61}  1 (1989).
\bibitem{Makar97}
A.S. Makarenko, M. Moskalkov and S. Levkov, {\em  Phys. Lett.,} {\bf  A 23} 391 (1997).
\bibitem{Kar03}
S. Kar, S.K. Banik and Sh Ray, {\em  Journ. of Physics A: Mathematical and Theoretical,} {\bf 36} 2271  (2003).
\bibitem{DDMSV}
 V.A. Danylenko, T.B. Danevych, O.S. Makarenko, S.I. Skurativs
kyi and V.A.
Vladimirov, {\em Self-Organization in Nonlocal Non-Equilibrium Media,} Subbotin Institute
of Geophysics, Kyiv, 2011.
\bibitem{Bril}
D.R. Brillinger, {\em  Mathematics Magazine,} {\bf 39} 145 (1966).
\bibitem{Evans_3}
J.W. Evans, {\em  Indiana Univ. Math. J.,} {\bf  22} 577 (1972).
\bibitem{Evans_4}
J.W. Evans, {\em Indiana Univ. Math. J.,}  {\bf  24} 1169 (1975).
\bibitem{Kapitula_Prom}
T. Kapitula, K. Promislov, {\em Spectral and dynamical stability of nonlinear waves}, Springer, New York, 2013.
\bibitem{Maurin}
K. Maurin, {\em Analysis, p. II}, PWN, Warsaw, 1972.
\bibitem{VladKu_04}Vladimirov V., Kutafina E., {\em Rep. Math.
Physics,} {\bf 54}, 261 (2004).
\bibitem{VladKu_05} Vladimirov V., Kutafina E., {\em Rep. Math.
Physics,} {\bf 56}, 421 (2005).
\bibitem{VladKu_06} Vladimirov V., Kutafina E., {\em Rep. Math.
Physics,} {\bf 58}, 465 (2006).
\bibitem{Vlad2008}
V. Vladimirov, {\em  Rep. Math. Physics,} {\bf 61} 380 (2008).
\bibitem{VlaMacz_12}
V.A. Vladimirov, Cz. M\c{a}czka,
{\em  Rep. Math.
Physics,}  {\bf 70} 313 (2012).







\end{thebibliography}
\end{document}